\documentclass[a4paper,11pt]{article}
\usepackage{pos}
\usepackage{url}
\usepackage{hyperref}
\usepackage{aas_macros}

\newcommand{\mmin}{m_{\rm min}}
\newcommand{\mmax}{m_{\rm max}}
\newcommand{\hu}{\,{\rm km \,s^{-1} \, Mpc^{-1}}} % Simone's comment

\title{Cosmology in the dark: How compact binaries formation impact the gravitational-waves cosmological measurements}
\author*[a]{S.~Mastrogiovanni}
\author[a]{K.~Leyde}
\author[b]{C.~Karathanasis}
\author[a]{E.~Chassande-Mottin}
\author[a]{D.~A.~Steer}
\author[c]{J.~Gair}
\author[d]{A.~Ghosh}
\author[e]{R.~Gray}
\author[f,g,h]{S.~Mukherjee}
\author[i,j]{S.~Rinaldi}

\affiliation[a]{Universit\'e de Paris, CNRS, AstroParticule et Cosmologie (APC), F-75013 Paris, France}

\affiliation[b]{Institut de F\'isica d'Altes Energies (IFAE), Barcelona Institute of Science and Technology, Barcelona, Spain}

\affiliation[c]{Max Planck Institute for Gravitational Physics (Albert Einstein Institute),
Am M\"uhlenberg 1, Potsdam 14476, Germany}

\affiliation[d]{Ghent University, Proeftuinstraat 86, 9000 Gent, Belgium}

\affiliation[e]{SUPA, University of Glasgow, Glasgow G12 8QQ, United Kingdom}

\affiliation[f]{Gravitation Astroparticle Physics Amsterdam (GRAPPA), Anton Pannekoek Institute for Astronomy and Institute for Physics, 
University of Amsterdam, Science Park 904, 1090 GL Amsterdam, The Netherlands}
\affiliation[g]{ Institute Lorentz, Leiden University, PO Box 9506, Leiden 2300 RA, The Netherlands}
\affiliation[h]{Delta Institute for Theoretical Physics, Science Park 904, 1090 GL Amsterdam, The Netherlands}
\affiliation[i]{Dipartimento di Fisica "E. Fermi", Università di Pisa, I-56127 Pisa, Italy}
\affiliation[j]{INFN, Sezione di Pisa, I-56127 Pisa, Italy}

\emailAdd{mastrosi@apc.in2p3.fr}

\abstract{Information about the mass spectrum of compact stars can be used to infer cosmological parameters from gravitational waves (GW) in the absence of redshift measurements obtained from electromagnetic (EM) observations. This method will be fundamental in measuring and testing cosmology with GWs for current and future ground-based GW detectors, since the majority of sources will be detected without an associated EM counterpart.
In this proceeding, we  discuss the prospects and limitations of this approach for studying cosmology. We  show that, even when assuming GW detectors with current sensitivities, the determination of the Hubble constant is strongly degenerate with the maximum mass for black hole production.
We discuss how assuming wrong models for the underlying population of black hole events can bias the Hubble constant estimate up to 40\%.}

\FullConference{%
  *** The European Physical Society Conference on High Energy Physics (EPS-HEP2021), ***\\
  *** 26-30 July 2021 ***\\
  *** Online conference, jointly organized by Universität Hamburg and the research center DESY ***
}

\begin{document}
\maketitle

\section{Introduction}

The detection first detection of gravitational waves (GWs) \cite{Abbott:2016blz, LIGOScientific:2018mvr} set a new milestone for cosmology, as GWs signals from compact binary coalescence directly provide a measure of the source luminosity distance.
If this measurement is supplemented with an evaluation of the source redshift, GW signals can be used to probe  the cosmic expansion history \cite{Sathyaprakash:2009xs, Holz:2005df}. This is of great interest given the current tension in the estimation of the Hubble constant $H_0$ today  \cite{Riess:2016jrr, Aghanim:2018eyx,Freedman:2017yms,Riess:2019cxk}.

The binary neutron star merger GW170817, and its electromagnetic (EM), was the first event considered for cosmological studies \cite{TheLIGOScientific:2017qsa} and provided a measurement of $H_0=70^{+19}_{-8} \hu$ \cite{Abbott:2017xzu} (68.3\% maximum posterior and highest density intervals). In this case, the redshift of the source was obtained from the observation of the GWs in association to a short gamma-ray burst with the subsequent kilonova and identification of host galaxy \cite{Chen:2020zoq,2021A&A...652A...1M}.

For this motivation, several methods that do not rely on a direct EM counterpart have been proposed. The first approach proposed in \cite{Schutz:1986,Holz:2005df} was to employ galaxy catalogs in correlation with the GWs localization volume to obtain a redshift estimation. This techniques provides a less accurate redshift estimation than a direct detection of the EM counterpart, but it can be applied to all the GWs sources, even binary black holes (BBHs) for which an EM counterpart is not expected. For instance, this technique was applied as case study to GW170817, obtaining $H_0={77}_{-18}^{+37} \hu$\cite{Fishbach:2018gjp}, but also to BBHs and neutron star black hole (NSBH) \cite{2021ApJ...915L...5A} detected during first three runs of the LIGO and Virgo detectors, see  \cite{LIGOScientific:2018mvr,Soares-Santos:2019irc,Abbott:2019yzh,2021JCAP...08..026F} for the different $H_0$ measures. However, even this method suffers from the limitation that \textit{(i)} galaxy catalogs only covers a limited fraction of the GWs sky localization and \textit{(ii)} galaxy catalogs rapidly becomes \textit{incomplete} in redshift \cite{Abbott:2019yzh}.

Here, we consider a different method, relying only on the GWs signal. The idea is to exploit the fact that as a consequence of the Universe expansion, the inferred binary masses are redshifted, i.e. $M_z= (1+z)M$, where $z$ is the redshift, $M$ is the mass the source and $M_z$ is the mass estimated at the detector. Hence, a redshift evaluation could be implicitly inferred from the knowledge, or using some assumptions on the source masses. This method was originally proposed in \cite{Taylor:2011fs, Taylor:2012db} for BNSs and extendend in more details in \cite{Farr:2019twy} for the mass spectrum of BBHs with Advanced LIGO and Virgo.

In this proceeding we focus on presenting in more details the impact of the BBHs source mass spectrum assumptions for the estimation. The remainder of the paper is organized as follows. In Sec.~\ref{sec:2} we resume the general method and its statistical implementation, in Sec.~\ref{sec:3} we highlight and discuss possible systematics in the $H_0$ and source mass reconstruction for BBHs observed with and without EM counterparts, while in Sec.~\ref{sec:4} we draw our conclusions and discuss future perspectives.

\section{Basics of the source mass method \label{sec:2}}

The idea is the following: given a set of set of $N_{\rm obs}$ GW detections with their estimations of luminosity distance $D_L$ and detector masses $m^{\rm det}_{1},m^{\rm det}_{2}$, one would like to reconstruct, for a given cosmology, the distribution of the binary redshift $z$ and source masses $m_{1},m_{2}$. 
This can be done by defining a \textit{hierarchical likelihood} \cite{Mandel:2018mve,2019PASA...36...10T,Vitale:2020aaz} that is used to calculate the probability to of a set of population parameters $\Lambda$,  that describes the source mass spectrum and redshift distribution of the binaries, together with cosmological parameters such as $H_0$. The posterior on the population level parameters and $H_0$ can be written as
\begin{equation}
    p(\Lambda, H_0|\{x\},N_{\rm obs}) \propto p(H_0,\Lambda) \prod_{i}^{N_{\rm obs}} \frac{\int p(x_i|H_0,z,m_1,m_2)p_{\rm pop}(z,m_1,m_2|\Lambda)dz dm_1 dm_2}{\int p_{\rm det}(z,m_1,m_2,H_0)p_{\rm pop}(z,m_1,m_2|\Lambda)dz dm_1 dm_2},
    \label{eq:posterior}
\end{equation}
where $p(H_0,\Lambda)$ is a prior term and $p(x_i|H_0,z,m_1,m_2)$ is the GW likelihood of the single event which is calculated from  the estimated $D_L, m^{\rm det}_{1},m^{\rm det}_{2}$ fixing a value of the $H_0$. The term $p_{\rm pop}(z,m_1,m_2|\Lambda)$ is prior term  describing the distribution of binaries in source mass and redshift that depends from the choice of several population parameters $\Lambda$. 
For instance, to describe the redshift distribution of the binaries, usually a powerlaw distribution in $(1+z)$ is implemented \cite{Abbott:2020gyp}, while for the source mass spectrum several phenomenological models are available. 
The denominator of Eq.~\ref{eq:posterior} takes into account selection biases thourgh the evaluation of a detection probability  $p_{\rm det}(z,m_1,m_2,\Lambda)$.

Roughly speaking, a value of $H_0$ will be preferred when the collection of reconstructed likelihoods in the source frame $p(x_i|H_0,z,m_1,m_2)$ matches the imposed population prior on $p_{\rm pop}(z,m_1,m_2|\Lambda)$. As an example, suppose the population prior $p_{\rm pop}(z,m_1,m_2|\Lambda)$ excludes  masses higher than $50 M_{\odot}$: if an event is found with masses $>50 M_{\odot}$ for a value of $H_0=30 \hu$ but not $H_0=70 \hu$, then the value of $H_0=30 \hu$ will be automatically excluded as it will return a null hierarchical likelihood.

Of course we do not know a priori the distribution of binaries in source mass and redshift and that is the motivation for which we need to jointly fit for population parameters related to the source mass spectrum and $H_0$ and other cosmological parameters.

\section{Results \label{sec:3}}

In order to quantify how the population assumptions impacts the estimation of $H_0$, in \cite{Mastrogiovanni:2021wsd} perform a simulation of BBHs detected by LIGO and Virgo detectors with sensitivities similar to the ones achieved during the second and third observing runs \cite{TheLIGOScientific:2014jea,TheVirgo:2014hva,Acernese:2019sbr,2020arXiv200801301I}.

In particular, we have chosen a uniform in comoving volume merger rate distribution in redshift. The source primary mass spectrum model is a linear combination of a decreasing powerlaw with slope $\alpha=2$ between minimum mass $\mmin=5 M_{\odot}$ and maximum mass $\mmax=85 M_{\odot}$, and a gaussian component with mean at $\mu_g=40 M_\odot$ and standard deviation of $\sigma_g=5 M_\odot$. We chose the fraction of events born in the gaussian peak to be $10\%$. For the secondary source mass, we chose a conditional powerlaw model\footnote{The condition being $m_2 \leq m_1$.}, with powerlaw slope $\beta=0$ (no preference for equal masses). This is a population model compatible with current O2 and O3 BBH events \cite{LIGOScientific:2018mvr,Abbott:2020niy}. The cosmology that we fix has  $H_0=67.7 \hu$ and $\Omega_{m,0}=0.308$ \cite{Ade:2015xua}.

We analyzed $N_{\rm inj} \leq 1024$ simulated events that passed a network signal-to-noise ration threshold of $12$. Below, we briefly discuss two cases from \cite{Mastrogiovanni:2021wsd}, the case in which no EM counterpart is observed and the case in which an EM counterpart is observed for all the events. We estimate jointly sampling from the posterior in Eq.~\ref{eq:posterior} calculated for the cosmological parameters $H_0, \Omega_{m,0}$ and the various population parameters.

\subsection{Dark sirens: Impact of the BBHs population assumptions}

We discuss in this section our findings when we assume that no EM counterpart is detected with the GW events.

Our first result concerns the impact of $\Omega_{m,0}$ on the determination of $H_0$. We perform two runs: in (i) we fix $\Omega_{m,0}$ to its injected value, while in (ii) $\Omega_{m,0}$ is able to vary in the range $[0.1,0.5]$ with a uniform prior. We find that for detector sensitivities comparable to O2 and O3, $\Omega_{m,0}$ only weakly impacts the determination of $H_0$. In fact, while fixing $\Omega_{m,0}$ we find that $H_0$ could be constrained to the 40\% accuracy (90\% credible intervals), while if we leave $\Omega_{m,0}$ able to vary, $H_0$ can be constrained at 50\% accuracy. See \cite{Mastrogiovanni:2021wsd} for more details.

\begin{figure}
    \centering
    \includegraphics[scale=0.35]{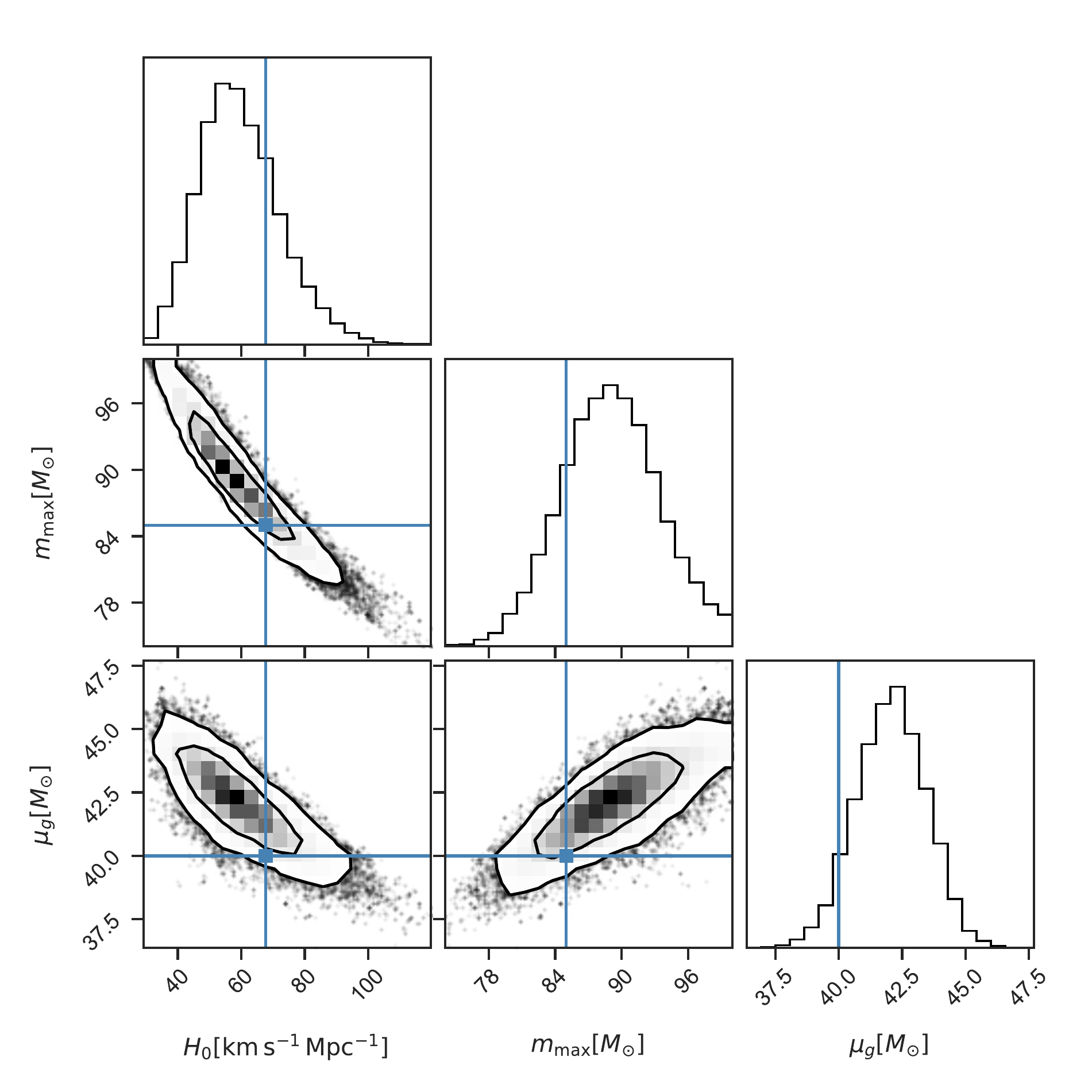}
    \includegraphics[scale=0.35]{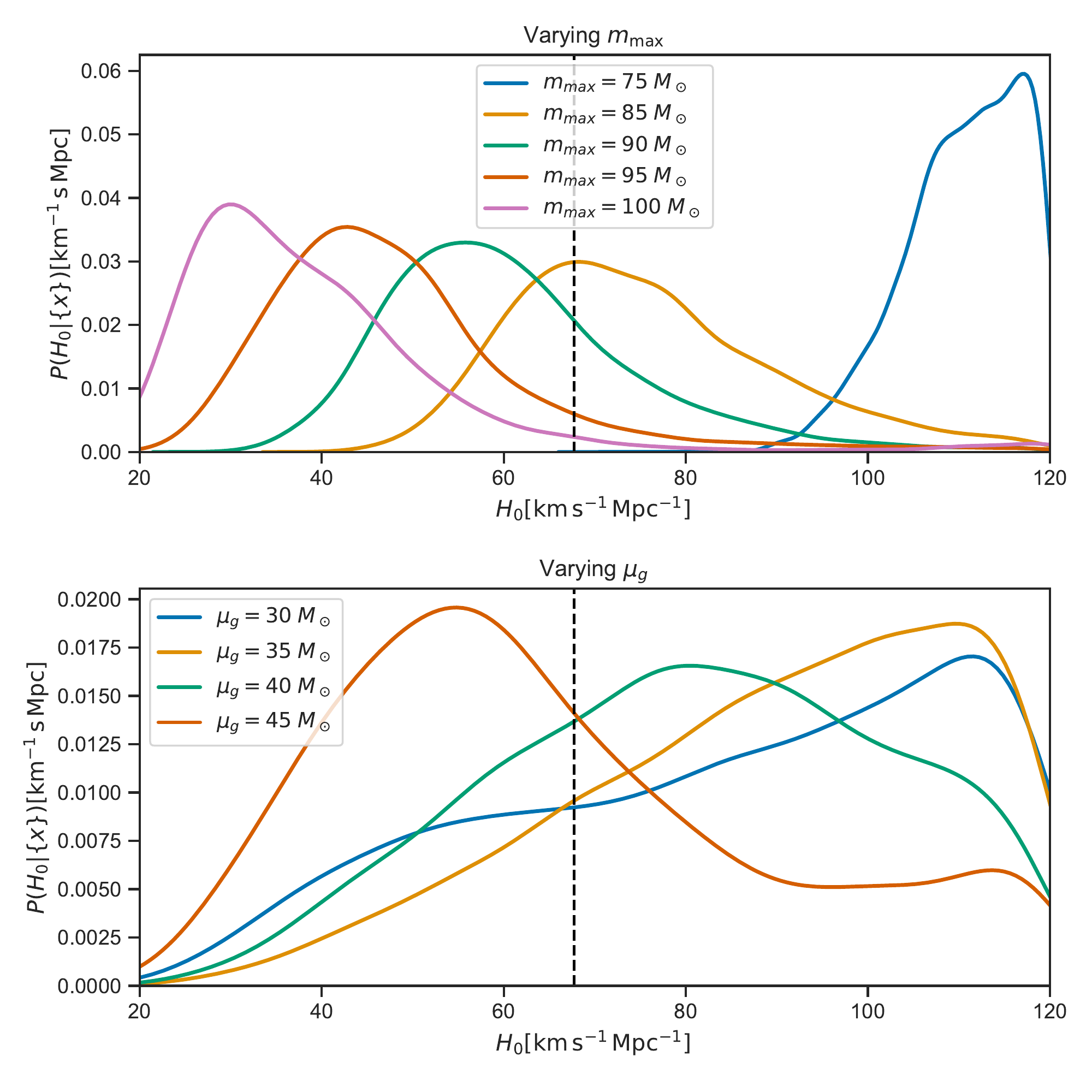}
    \caption{\textit{Left}: Posterior distribution on the $H_0$, $\mmax$ and $\mu_g$ for 64 BBH events detected with LIGO and Virgo at current sensitivities. The blue lines show the true parameters. The contours indicate the $1\sigma$ and $2\sigma$ confidence level intervals. \textit{Right:} Posterior distribution for $H_0$ obtained by fixing $\mmax$ and $\mu_g$ in a range around their true values  $ \mmax = 85 M_{\odot}$ and $\mu_g = 40 M_{\odot}$. The black dashed line indicates the true value of $H_0$. Figure from \cite{Mastrogiovanni:2021wsd}.}
    \label{fig:H0_mmax}
\end{figure}

Our second result concerns the interplay between the determination of $H_0$ and other population parameters, for the simulated population $\mmax$ and $\mu_g$.
In Fig.~\ref{fig:H0_mmax} left panel, we show the joint posterior distribution between $H_0, \mu_g$ and $\mmax$ for 64 BBHs without EM counterpart.

Indeed, $\mmax$ and $\mu_g$ impact the estimation of the $H_0$. Let us take $\mmax$ as an example: lowering $H_0$ values move the observed GW source to lower redshifts (remember that what it is estimated is actually $D_L$), as a consequence, this will push the source masses to higher values (since we measure detector masses). Hence, to ``correct'' for this increase in source masses, an higher $\mmax$ is needed and this creates a correlation between the determination of $H_0$ and $\mmax$. A similar discussion is valid for the position of the gaussian peak $\mu_g$.

The interplay, or correlation between the determination of $H_0$ and population parameters related to the source mass spectrum, can also become a source a bias for the determination of the $H_0$. In fact, if an anlysis is performed by fixing a erroneous population model for BBHs, then the estimation of $H_0$ might be systematically biased.

As an example, in Fig.~\ref{fig:H0_mmax} right panel where we show posterior distributions obtained for $H_0$ when fixing either $\mu_g$ or $\mmax$ mismatched values with respect to the injected ones. 
One can see that already with 64 events, the amount of bias due to wrong population assumptions is significant. We therefore argue that population assumptions and cosmological parameters estimation should be taken under control together.

\subsection{Bright sirens: Impact of the populations assumptions}

We also considered a different situation, namely \textit{What if BBHs are provided with EM counterparts? Will the source mass spectrum assumptions be important?}. This is of course an unrealistic situation but it will give us an idea on the interplay of population assumptions and EM counterparts in this case.

In order to take into account the extra redshift estimation $z_{\rm obs}$ from the EM counterpart, we have to modify Eq.~\ref{eq:posterior}. We obtain, with the assumption that the GW likelihood is separable in a term dependent from the luminosity distance $p(x^i|H_0,z^{i}_{\rm obs})$ and in one from the masses $p(x^i|m_1,m_2,z_{\rm obs})$,
\begin{equation}
     \frac{p(\Lambda|\{x\},z_{\rm obs})}{p(\Lambda)}  \propto   \prod_i^{N_{\rm obs}}  \frac{\int p(z^{i}_{\rm obs}|\Lambda) p(x^i|H_0,z^{i}_{\rm obs})p(x^i|m_1,m_2,z_{\rm obs}) p_{\rm pop}(m_{1},m_{2}|\Lambda) dm_{1} dm_{2}}{\int p_{\rm det}^{\rm GW}(m_1,m_2,z,H_0)p_{\rm det}^{\rm EM}(m_1,m_2,z,H_0)\:p_{\rm pop}(m_1,m_2,z|\Lambda)dm_{1} dm_{2} dz}.
     \label{eq:EM_simp_ter}
\end{equation}
Above, $p_{\rm det}^{\rm EM}$ and $p_{\rm det}^{\rm GW}$ represent the probability of detecting an EM counterpart and a GW given a set of binary parameters.
It is important to notice that the two last terms depend individually on either the population or cosmological parameters, while $p(z^{i}_{\rm obs}|\Lambda)$. 

From Eq.\ref{eq:EM_simp_ter} we can see that if we fix $\Lambda$ to incorrect values, the evaluation of  Eq.~(\ref{eq:EM_simp_ter}) will only differ from a normalization constant and $H_0$ will not be impacted. Of course, the calculation of selection effects in the denominator should be calculated correctly and the population source spectrum assume should include the binary masses in its spectrum.

\begin{figure}
    \centering
    \includegraphics[scale=0.5]{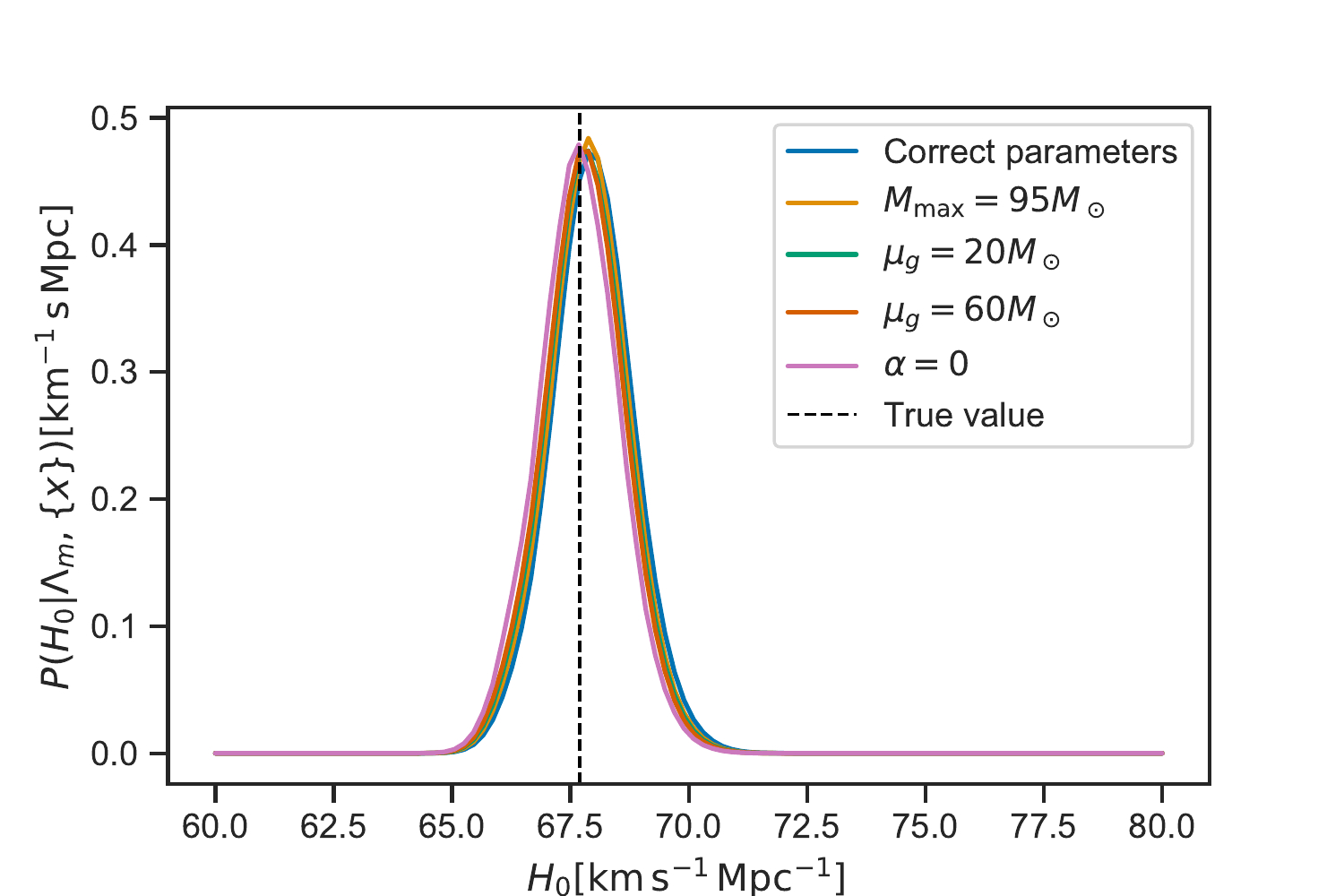}
    \caption{Hubble constant posterior generated from 64 our synthetic population of BBHs fixing different population models and providing the redshift of the GW source (assumed from an EM counterpart). An incorrect choice of one of the population parameters (see legend) does not affect significantly the $H_0$ estimation. The vertical dashed line indicates the injected value. Note that for this plot the posterior samples are generated taking into account the correlations between masses and luminosity distance (point (ii) above is dropped). Figure from \cite{Mastrogiovanni:2021wsd}}
    \label{fig:em_counterpart}
\end{figure}

In Fig.~\ref{fig:em_counterpart} we show an $H_0$ posterior computed with 64 events of the simulated population and assuming an EM counterpart. One can see that wrong population assumptions are not particularly important.

\section{Conclusions \label{sec:4}}

In this paper we have shown the interplay between BBHs population assumptions and the estimation of cosmological parameters with GWs standard sirens.
We have shown that parameters that governs features in the source mass spectra of BBHs are likely to introduce strong correlations with the determination of the cosmological parameters, in particular $H_0$.
On one hand, the source mass spectrum provides a useful channel to investigate cosmology with GWs even in absence of EM counterparts or with incomplete galaxy catalogs. On the other hand, if the source mass spectrum is mismatched, a significant bias on the evaluation of $H_0$ is introduced.

We have also shown that in presence of an EM counterpart, the dependence of the $H_0$ estimation from population assumption can be quenched.

In the future, other interesting methods  for instance the cross-correlating  GW sources with galaxies, see \cite{Oguri:2016dgk,Mukherjee:2019wcg, Mukherjee:2020hyn} can also help leverage the determination of cosmological parameters with GWs.

\acknowledgments
SM is supported by the LabEx UnivEarthS (ANR-10-LABX-0023 and ANR-18-IDEX-0001), of the European Gravitational Observatory and of the Paris Center for Cosmological Physics. KL is grateful to the Fondation CFM pour la Recherche in France for supporting his PhD.
The authors are grateful for computational resources provided by the LIGO Laboratory and supported by National Science Foundation Grants PHY-0757058 and PHY-0823459. CK is partially  supported   by  the Spanish MINECO   under the grants SEV-2016-0588 and PGC2018-101858-B-I00, some of which include ERDF  funds  from  the  European  Union. IFAE  is  partially funded by the CERCA program of the Generalitat de Catalunya. RG is supported by the Science and Technology Facilities Council. SMu is supported by the Delta ITP consortium, a program of the Netherlands Organisation for Scientific Research (NWO) that is funded by the Dutch Ministry of Education, Culture, and Science (OCW). LIGO is funded by the U.S. National Science Foundation. Virgo is funded by the French Centre National de Recherche Scientifique (CNRS), the Italian Istituto Nazionale della Fisica Nucleare (INFN), and the Dutch Nikhef, with contributions by Polish and Hungarian institutes. This material is based upon work supported by NSF’s LIGO Laboratory which is a major facility fully funded by the National Science Foundation.

\bibliographystyle{JHEP}
\bibliography{refs}

\end{document}